\documentclass[aps,pra,twocolumn,groupedaddress]{revtex4-1}

\usepackage{graphics}
\usepackage{draftcopy}

\begin{document}

\title{Laser frequency offset locking scheme for high-field imaging of cold atoms}


\author{Graciana Puentes}
\affiliation{MIT-Harvard Center for Ultracold Atoms, Research Laboratory of Electronics, and Department of Physics, Massachusetts Institute of Technology, Cambridge, Massachusetts 02139, USA, \email{gpuentes@mit.edu}}


\date{\today}

\begin{abstract}
We present a simple and flexible frequency offset locking scheme developed for high-field imaging of ultra-cold atoms which relies on commercially available RF electronics only. The main new ingredient is the use of the sharp amplitude response of a ``home-made" RF filter to provide an error signal for locking the lasers. We were able to offset lock two independent diode lasers within a capture range of 200 MHz, and with a tuning range of up to 1.4GHz. The beat-note residual fluctuations for offset locked lasers are bellow 2MHz for integration times of several hundreds of seconds. 
\end{abstract}

\pacs{9908098908}

\maketitle

\section{Introduction}
\noindent When absolute laser frequency stabilization is required it is customary to lock the frequency of the laser to an atomic standard, such as in saturated absorption spectroscopy, or alternatively to a master laser operating at a known frequency via an optical phase-locked loop [1]. However, when phase coherence between two lasers is not required there are simpler locking methods which have lower bandwidth requirements, such as frequency offset locking. \\

\noindent Fequency offset locking schemes work by interfering two lasers at a beam-splitter and detecting the beat-note $\omega_{\mathrm{BN}}$  with a fast photo-diode. By locking $\omega_{BN}$ to a constant fixed value an active feedback system ensures a fixed offset between the master and slave lasers. It is then possible to offset the two lasers by an arbitrary amount (within the scope of the given locking scheme) by mixing the beat signal with an external local oscillator (LO) at a given known frequency ($\omega_{\mathrm{LO}}$). In order to compensate for frequency drifts and stabilize the two lasers to a constant offset an error signal is generated and a feedback correction is supplied to the slave laser by means of a servo controller loop.\\

\noindent The main distinction between different offset locking schemes is due to the method for generating the error signal to servo-lock the slave laser to the master laser.~In References~[2,3], the authors present a method which utilizes an electronic delay line in order to produce a frequency dependent phase shift. However, a drawback of such a scheme is the limited capture range due to the presence of several zero-crossings in the error signal, which result in multiple lock points. Another possibility for generating the error signal is to use a commercial frequency-to-voltage converter [5], though such converters are not usually available at high operating frequencies and their cost can be a limiting factor.~Here we present a very flexible and affordable scheme for frequency offset locking which is based on generating the error signal from the amplitude response of a home-built RF electronic filter. The advantage of this scheme is that the error signal response is linear (within the working range of the electronic components) and thus provides for a unique zero point for offset locking the lasers in an arbitrary frequency range (i.e. up to several GHz) only limited by the bandwidth of the electronic components.~Furthermore, the scheme is based on commercially available electronics only and thus simplifies the design requirements of Ref. [4], while still providing for a similar performance, as quantified for instance by the capture range.~We expect such simple and efficient frequency offset locking scheme to be of interest to the ultra-cold atom community.  \\

\section{Applications: High field imaging}

\noindent The ability to image clouds of ultra-cold atoms at high magnetic fields has several advantages. Switching off large magnetic fields takes a characteristic time of hundreds of  $\mu$s due to the  inductance of the magnetic coils and thus such a measurement can not be considered instantaneous. This would be an impediment for instance in the study of non-equilibrium dynamics. Furthermore, in strongly interacting systems other relevant times scales, such as the coherence time of a superfluid state, can be even shorter than hundreds of $\mu$s, and the long range coherence might have decayed after the fields are switched off without allowing enough time for imaging and characterization. However, high magnetic fields can shift the resonant wavelength of the atomic transition to the point where the lasers used for trapping and cooling cannot be used for imaging. In particular, the presence of external magnetic fields modifies the energy sub-levels of the two magnetic moments ($\vec{I},\vec{J}$) coupled by the hyperfine interaction in a non-linear manner as given by the Breit-Rabi formula [6]. In the limit of strong fields the hyperfine interaction ($\vec{I} \cdot \vec{J}$) is essentially decoupled. In particular, the hyperfine splitting of the ground state of ${}^{7}\mathrm{Li}$ ($2 {S}^2_{1/2}$) is subdivided into two manifolds, upper and lower, corresponding to $m_{J}=1/2$  and $m_{J}=-1/2$ (high field seeking states), respectively. In Fig. 1 we show the calculated Breit-Rabi splitting due to the presence of an external magnetic field in the range of 0 to 1000G, a range in which ${}^{7}\mathrm{Li}$ shows broad Feshbach resonances. The two manifolds contain four sub-levels corresponding to $m_{J}=\pm 1/2$ and  $|m_{I}|\leq 3/2$ ( F=2 and F=1 at B=0). The energy splitting between the upper and lower manifold is of the order of 1.4 GHz for a magnetic field in the range of 700G. These values are entirely compatible with the tuning range of the offset locking scheme presented here, where the main limiting factor is given by the bandwidth of the fast photo-diode and amplifiers and which can be easily increased by using faster detectors with suitable amplification. 

\begin{figure} \resizebox{0.9\hsize}{!}{\includegraphics*{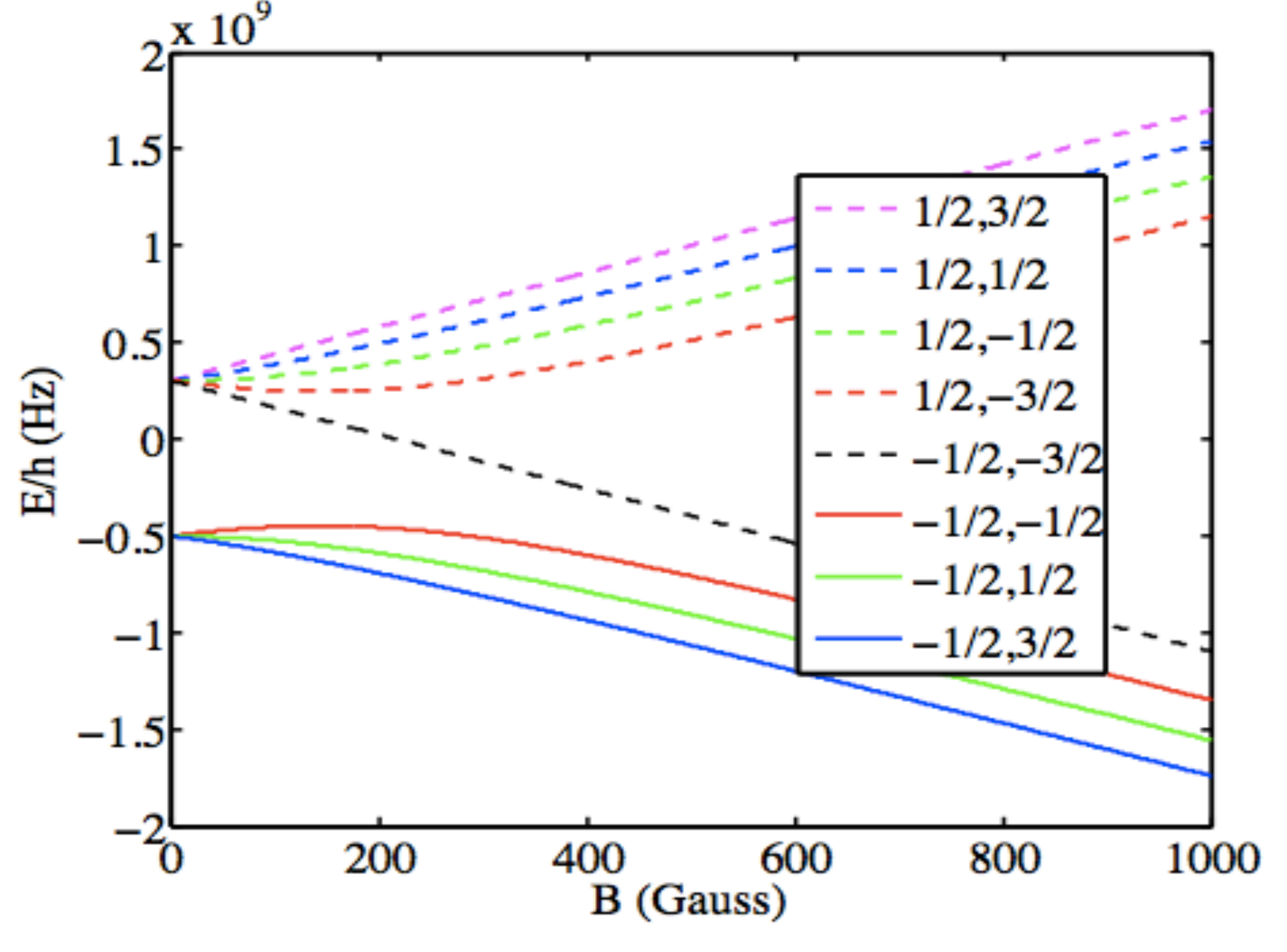}} \caption{Breit-Rabi diagram in ${}^{7}\mathrm{Li}$ for relevant magnetic field strengths (B). At large external magnetic fields the hyperfine interaction is decoupled and there are two manifolds (upper and lower) corresponding to $m_{J}=\pm1/2$ indicated by dashed and solid lines, respectively. The  energy levels are indicated by quantum numbers $(m_{J},m_{I})$. The energy splitting between the upper and lower manifold is within 1.4 GHz for a magnetic field in the range of 700G. These values are entirely compatible with the offset locking scheme presented here. }
\end{figure}

\section{Offset lock scheme}

\noindent The frequency offset locking scheme  is shown in Fig. 2. The amplified output beat-note ($\omega_{\mathrm{BN}}$) of a fiber-coupled fast photo-diode (PD Model: Thorlabs SIR5-FC, bandwidth 2 GHz) is mixed with the RF signal produced by a function generator (Model: SRS SG384, bandwidth 4 GHz), at an RF frequency mixer (Model: Mini-circuits ZFM-11-5+). In our implementation we used an RF amplifier (LNA-1400 RF Bay) which is suitable for the input-output signal levels required in the circuit. Note that the amplifier bandwidth is 1.4GHz, which sets an upper limit to the tuning range for the $\omega_{\mathrm{BN}}$ that can be processed by this circuit. In order to test the level of the amplified beat-note signal we introduced an RF coupler (CPL1 Model: ZFDC-10-5-S+ Mini-circuits). The signal after the mixer is the convolution of the two incoming frequencies $\omega_{\mathrm{BN}}$ and $\omega_{\mathrm{LO}}$, in such way that we should obtain both the frequency sum and frequency difference $\omega_{+}=\omega_{\mathrm{BN}}+\omega_{\mathrm{LO}}$ and $\omega_{-}=|\omega_{\mathrm{BN}}-\omega_{\mathrm{LO}}|$. In order to work only with the frequency difference $\omega_{-}$, we use a low pass filter (LP150,  Model: SLP-150+ Mini-circuits )  which is in the range of the frequency response of our RF filter. The mixed signal after LP150 is also monitored via an output coupler CPL2 at an RF spectrum analyzer and, before locking, it should show a single peak centered at $\omega_{-}=|\omega_{\mathrm{BN}}-\omega_{\mathrm{LO}}|$. This frequency difference will eventually be locked to a constant value using the error signal given by the filter response. \\

\begin{figure} \hspace{-0.5cm} \resizebox{1.0\hsize}{!}{\includegraphics*{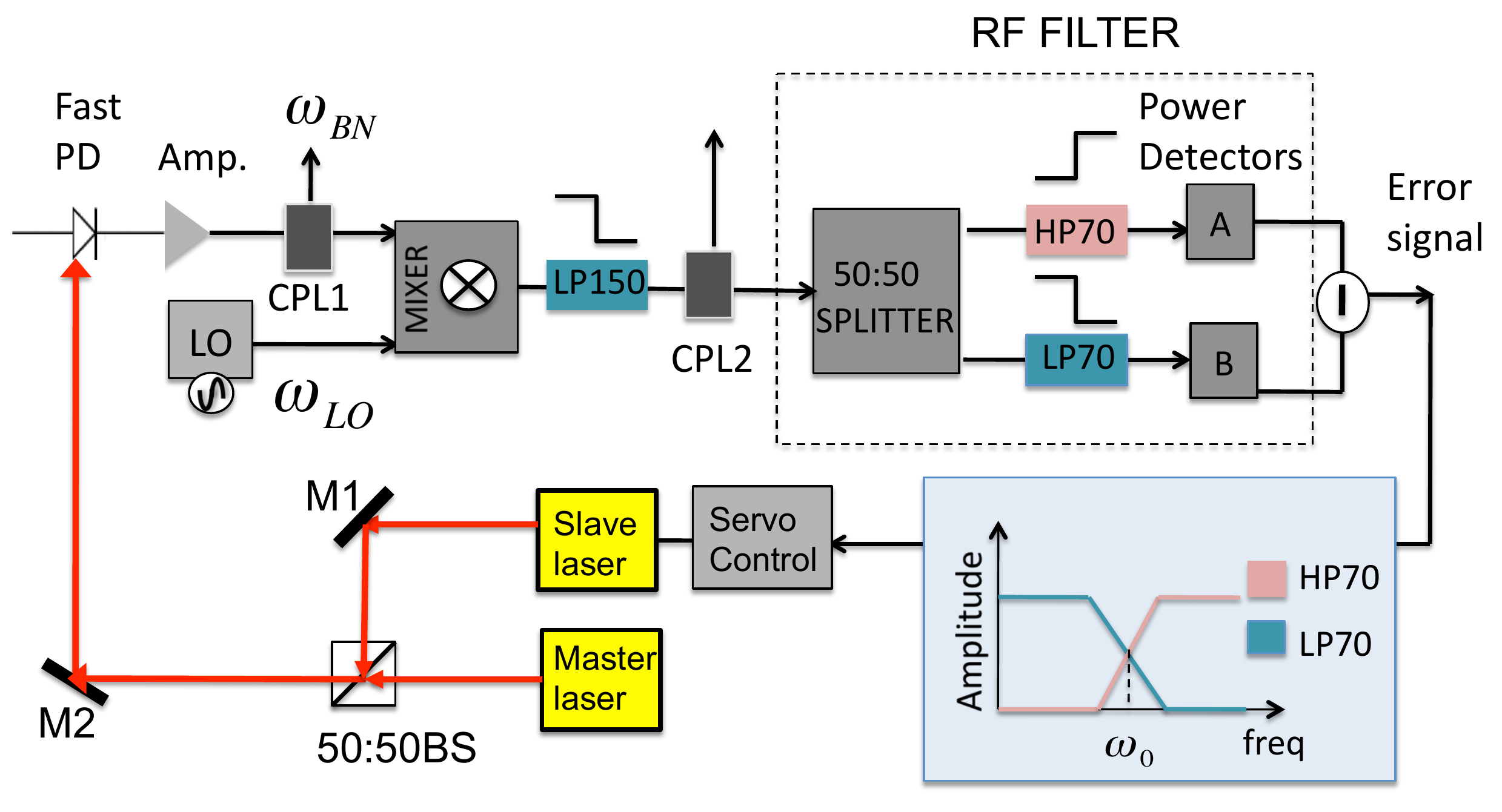}} \caption{Offset locking scheme using a ``home-made" RF filter. The inset shows a graph of the comercial high-pass filter  and low-pass filter amplitude response used to produce the error signal.}
\end{figure}

\noindent The RF filter is composed of a 50:50 splitter (Model: ZMSC-2-1+ Mini-circuits), a low pass filter (LP70, Model: SLP-70+ Mini-circuits) and a high pass filer (HP70 Model: SHP-100+ Mini-circuits), and two power detectors A and B (Model: ZX47-40LN-S+ Mini-circuits) which provide the different responses for the error signal. The responses of the filters LP70 and HP70 are pictured on the insets of Fig. 2. These filters were chosen because they have a linear decreasing and increasing response, respectively, in the vicinity of  the target range of $\omega_{-}$. Since the response of the two detectors is monotonic, and with opposite slopes, there is a single crossing which determines the fixed value $\omega_{0}$ to which the scheme will lock the signal, which is  around 70MHz. As will be explained below the exact value of $\omega_0$ depends on the input RF power. When $\omega_{-}=\omega_{0}$ detector A and detector B will detect the same signal and therefore there will be no difference (or error) to correct for. Next, when $\omega_{-}>\omega_{0}$ the response from  A will be larger than the response from B and therefore the difference between A and B  (error signal) will be positive. On the other hand, when $\omega_{-}<\omega_{0}$ the response from A is lower than that of B and the error signal will be negative. In this way, we have produced an error signal which has a linear dependence on frequency with which to lock our slave laser. In particular, it should be noted that this locking scheme sets $\omega_{-}=\omega_{0}$. We note that there are two  dualities with respect to the locking point. One is due to the actual value of the beat-note, since the sign of $\omega_{BN}$ is not known \emph{a priori}. Nevertheless, the two possible beat-notes are sufficiently far away from each other and should not introduce additional locking points. The second ambiguity  is due to the actual sign of $\omega_{-}=|\omega_{BN} - \omega_{\mathrm{LO}}|$, which means that the slave laser can be offset locked to either $\omega_{\mathrm{BN}}=\omega_{\mathrm{LO}}+/-\omega_{0}$. Depending on the LO frequency ranges, this duality can introduce a second zero crossing to lock the lasers which should be offset by roughly 140MHz. The error signal is fed to a servo controller which supplies a linear feedback to the slave laser current and thus warrants that the two lasers remain at a constant offset, at the expense of modifying the beat-note frequency.

\section{Experimental results}

\noindent Two independent (master and slave) diode lasers (Model: Toptica DL100 and Toptica TApro) centered  at 671 nm were used to implement the scheme shown in Fig. 2. The line-width of the lasers is typically of the order of  a few hundred kHz and the mode-hop-free region is several GHz wide. The master laser will eventually be locked to the D2 line of ${}^{7}\mathrm{Li}$ by means of saturated absorption spectroscopy, but it does not require to be locked for the  purpose of this (proof of principle) paper.~The lasers are coherently combined at a fiber beam-splitter, and from this interference signal we obtain the beat-note. Care should be taken that the two lasers are within a few hundred MHz separation at the start. In order to   control the initial frequency difference we monitored both lasers on a high finesse wavelength meter (WS/7 Angstrom). This beat-note is first amplified and then mixed with the LO at a frequency mixer, as described in the previous section. We observe both the beat-note $\omega_{BN}$ and the frequency difference $\omega_{-}$ at two consecutive output couplers CPL1 and CPL2, respectively (see Fig. 2). \\

\noindent The experimental results are displayed on Fig.3 to  Fig.6. The input optical power was 540$\mu$W, with equal contributions in both laser modes. The RF spectral traces at different stages of the locking scheme were obtained on a spectrum analyzer display (R3463 Advantest modulation). Fig. 3 shows an RF spectral trace of the beat-note and local oscillator measured at the first output coupler (CPL1) and observed on the display of the spectrum analyzer. The value in this case is $\omega_{BN}=325$MHz with a full width at half maximum (FWHM) of 10MHz. We chose this particular value since the difference with a local oscillator centered around $400$MHz gives a frequency in the vicinity of the filter characteristic frequency $\omega_0$. The attenuation per division in the spectral traces is 10dB. \\

\begin{figure} \hspace{-0.5cm} \resizebox{0.95\hsize}{!}{\includegraphics*{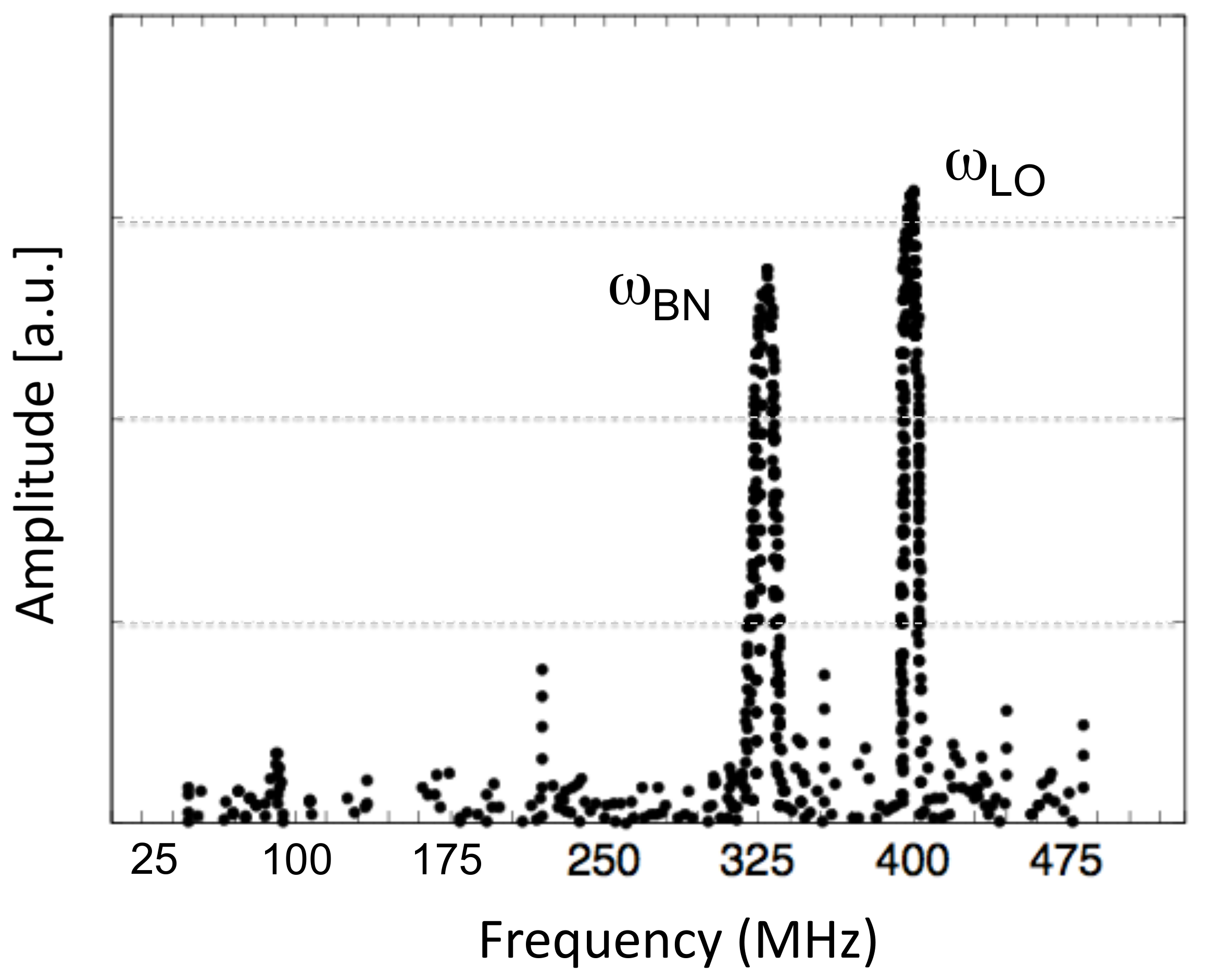}} \caption{ Beat-note (BN) and local oscillator (LO) spectral traces displayed on same scale. The beat-note was chosen to be $\omega_{BN}=325$MHz and the LO frequency was  $\omega_{LO}=400$ MHz. The attenuation/div is 10dB. }
\end{figure}

\noindent In order to test the locking capacity of the scheme, as well as the range of LO frequencies ($\omega_{LO}$) and beat-note frequencies ($\omega_{BN}$) for which the two lasers remain locked, the so called capture range, we measured the beat-note for different LO frequencies on a spectrum analyzer  both for the locked  and unlocked case. In Fig. 4 we show the measured $\omega_{BN}$ for different LO frequencies in the range of 200MHz to 600MHz both when the lasers are locked (solid line) and unlocked (dashed line).~As expected, for the unlocked case  $\omega_{BN}$ is independent of $\omega_{LO}$, and fluctuates around a nominal value of 325 MHz, due to laser frequency drift. On the contrary, when the two lasers are offset locked there is a linear dependance between the LO and beat-note frequency due to the offset constraint $\omega_{-}=|\omega_{BN}-\omega_{LO}|$. We note that the capture range is of the order of 200MHz, after which the lasers drop out of lock. This range is set by the low pass filter LP150 (see Fig. 2), and can in principle be increased arbitrarily by using a broader filter. For LO frequencies higher than 400 MHz the lasers are unlocked, and we note the appearance of a mode-hop for an LO frequency in the range of 500 MHz. This might indicate that the feedback loop tries to lock the laser to the next locking point, given by the duality in $\omega_{-}=\omega_{BN}-\omega_{LO}$ and $\omega_{-}=\omega_{LO}-\omega_{BN}$.

\hspace{-0.6cm}
\begin{figure} \resizebox{0.95\hsize}{!}{\includegraphics*{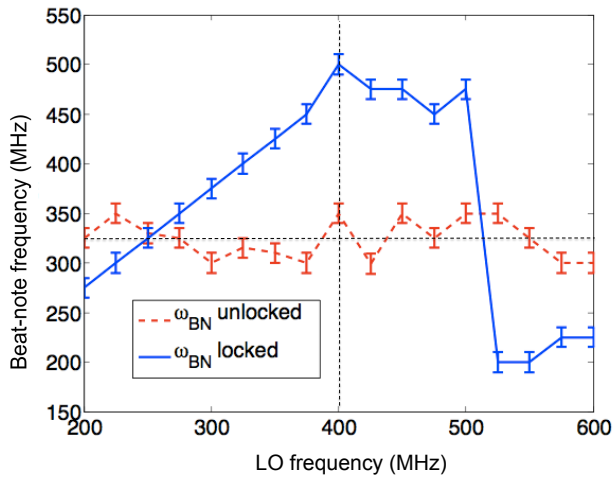}} \caption{Measured beat-note frequency $\omega_{BN}$ (in MHz) for different LO frequencies $\omega_{LO}$. When the lasers are not offset locked (dashed line) there is no correlation between these two frequencies, and $\omega_{BN}$ remains in the vicinity of 325MHz independently of $\omega_{LO}$. When the lasers are offset locked (solid line) there is a linear dependence between $\omega_{BN}$ and $\omega_{LO}$ due to the offset constraint $\omega_{0}=|\omega_{BN}-\omega_{LO}|$. In this figure we note that the capture range is of the order of 200MHz, while for LO frequencies above 400MHz  the lasers drop out of lock.~The noise in the figure is due to laser frequency drift. }
\end{figure}


\noindent The error signal (in Volts) is obtained by means of a servo controller (Model New Focus S100), by subtracting to measured powers in detectors A and B (see Fig. 2) and providing a linear feedback on the current of the slave diode laser. Better stability can be obtained by feeding back onto the laser piezo, though for this first demonstration we found that a feedback on the current was enough. The error signal should be within the voltage range of the diode laser and is adjusted by setting the gain of the servo lock box. The output offset in the servo controller is set to zero to ensure that the lasers are locked to the zero crossing of the error signal. When the  lasers are offset locked the signal fluctuations over time  are below 100mV. \\

\noindent Fig. 5 shows the error signal obtained by subtracting the output power in detectors A and B at the servo controller for different LO frequencies in the range of 250MHz to 400MHz. Changing the LO frequency modifies $\omega_{-}$ and this produces a different error to correct for. We note that there is a unique zero crossing within a range of roughly  140MHz due to the duality sign in the frequency difference and that in a range of roughly 10MHz around the locking point the response of the filter is linear and with a steep slope, as expected.\\

\begin{figure} \resizebox{0.95\hsize}{!}{\includegraphics*{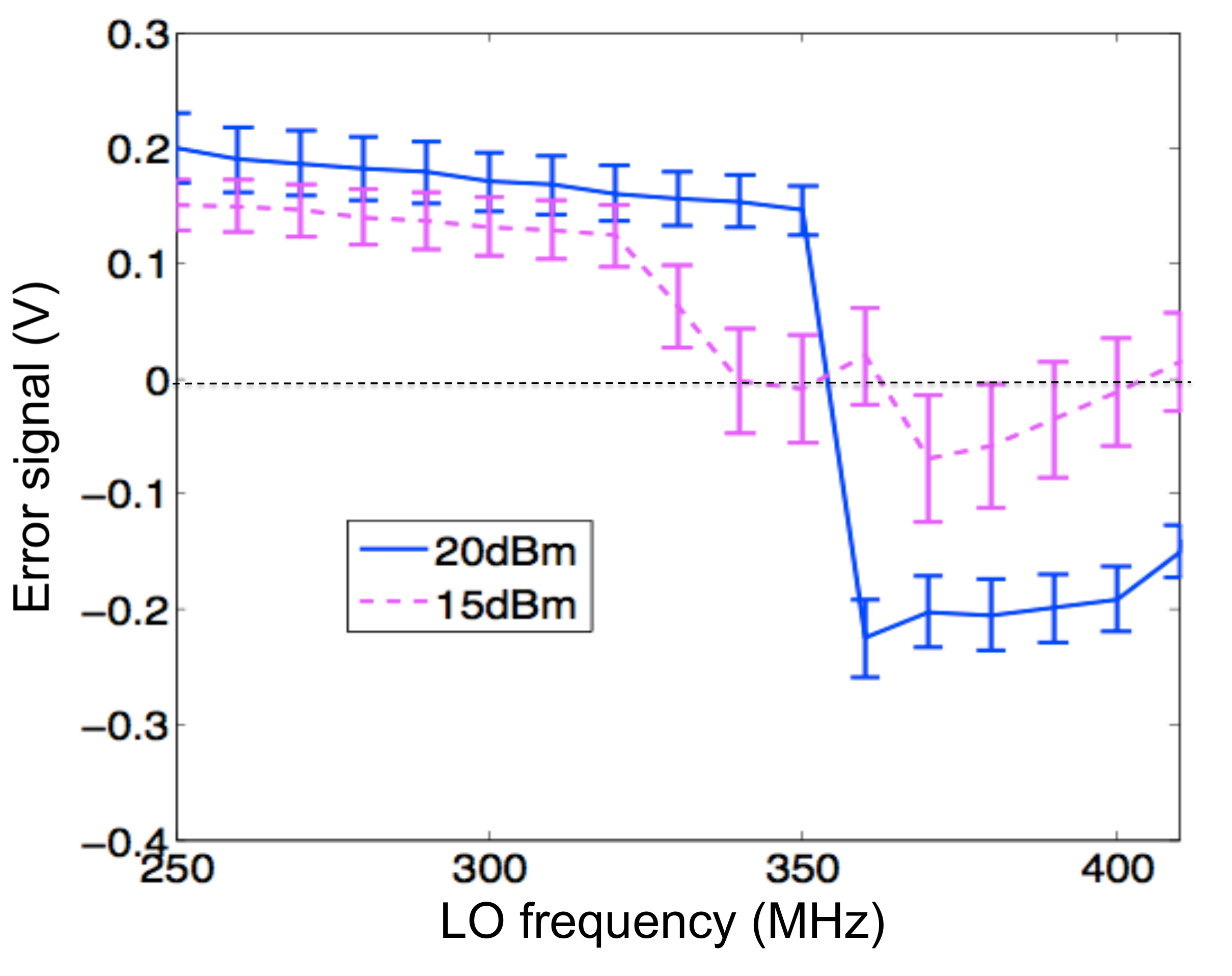}} \caption{Measured error signal as a function of LO frequency showing the sharp linear response of the RF filter. The zero of this error is used to offset lock the lasers. The power level in the beat-note was tuned  to 20dBm and 15dBm by changing the input optical power. For low input RF powers there is a clear degradation of the performance of the filter mainly due to low mixer performance. }
\end{figure}

\noindent Next, we measured the error signal for different input powers in the beat-note of 15dBm and 20dBm. The different powers were obtained by changing the input optical power. Another possibility is to use RF attenuators in the circuit. As expected the zero crossing depends on the input power, due to the dependance of the response of the mixer and filters on the input power level.~We note that the performance of the filter is significantly degraded for input  powers below 15dBm. \\


\noindent Finally, we characterized the frequency stability of the offset locked beat-note ($\omega_{BN}$) over time by sampling $\omega_{BN}$ at a repetition rate of Hz over an integration time of 325s, this is shown in Fig. 6. When the two lasers are not offset locked (Fig. 6 (b), dashed line)  the beat-note fluctuates over time within a range of roughly 20MHz, for an integration time of 325s. While, for offset locked lasers (Fig. 6 (a), solid line) frequency fluctuations are within a few MHz. The standard deviation ($\sigma$) of each sequence is $\sigma_{a}=0.54$ and $\sigma_{b}=5.54$, for offset locked and unlocked lasers, respectively, showing a 10-fold reduction in $\sigma$ for the offset locked case. The inset shows an histogram of the frequencies during the integration time, showing a discontinuous drift over tens of MHz in the unlocked case (b), and a continuous drift (no mode-hops) within less than 2 MHz for the locked case (a). Higher frequency stabilities can be obtained by feeding back on the slave laser piezo.  \\

\section{Considerations}

\noindent Some considerations for the optimal performance of the scheme are required. 
First, both the master and slave lasers should be as single mode as possible so that the beat-note can be  characterized by a sharp peak in the spectrum analyzer (FWHM 10MHz) within a sufficiently large offset frequency range (at least 1GHz). Mode hopping will broaden the beat-note and reduce the performance of this locking scheme. \\


\noindent Next, the total optical power in the beat-note should be in the range of 500$\mu$W, with both lasers having similar contributions in order to increase the visibility in the beat-note. Such input power  should produce a signal at the first output coupler CPL1 of at least 20dBm (see Fig. 5). Also, when interfering the two relevant lasers it is convenient to use a fiber beam splitter to obtain a cleaner beat note. \\

\begin{figure}  \hspace{-1.0cm} \resizebox{1.1\hsize}{!}{\includegraphics*{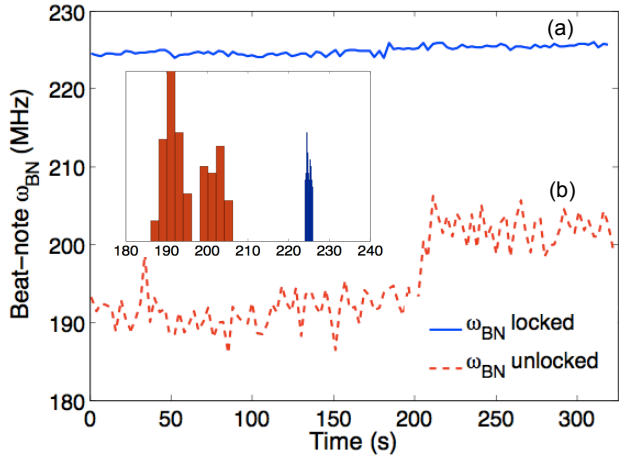}} \caption{(Color online) Beat-note signal vs time for  (a)  offset locked lasers and (b) unlocked lasers. The frequency fluctuations during an integration time of 325s  are (a) within 2MHz and (b) within 20MHz. The standard deviation ($\sigma$) of each sequence are (a) $\sigma_{a}$=0.54 and (b) $\sigma_{b}$= 5.54, respectively. Inset shows an histogram of frequencies in MHz.  }
\end{figure}

\noindent The LO frequency should be sufficiently large ($>$ 200MHz) for the low-pass filter LP150 to filter the sum frequency $\omega_{BN}+\omega_{LO}$. The lasers should be sufficiently close so that the difference between the LO frequency and the beat-note signal is well bellow 150MHz. As mentioned previously, the capture range of the locking scheme can be arbitrarily  increased by choosing a broader low-pass filter. \\


\noindent Finally, a clear signature of the good performance of the offset locking scheme can be obtained by observing the difference frequency $|\omega_{BN}-\omega_{LO}|$ on the second output coupler. When the lasers are not locked this difference should be arbitrary and below 150MHz. While this difference should be fixed at the characteristic filter frequency $\omega_0$ when the two lasers are locked. Note that the characteristic frequency $\omega_0$ is in the range of 70MHz, but the specific value depends on the RF input power. In this implementation we have $\omega_0=67.96$ MHz.




\section{Conclusions}

\noindent We have successfully demonstrated a very simple and flexible scheme for laser frequency offset locking based on the linear response of a ``home-made"  filter which relies on commercially available RF electronics only and therefore simplifies the design requirements of previous schemes [4], while providing for a similar locking performance. Two independent diode lasers were offset locked within a capture range of several hundreds of MHz. The capture range of this scheme is only limited by the bandwidth of the filters and can be increased arbitrarily. The tuning range, in turn, is 1.4GHz and is only limited by the amplifier bandwidth. We plan to use the current scheme for high field imaging of ultra-cold ${}^{7}\mathrm{Li}$. \\

\section{Acknowledgements}

\noindent The author acknowledges Wolfgang Ketterle, Aviv Keshet, Edward Su, and Christian Sanner for useful discussions and technical support.~The author is thankful to David Weld for critical reading of the manuscript. This work was supported by the NSF through the Center of Ultracold Atoms, by NSF award PHY-0969731, through an AFOSR MURI program, and under ARO Grant No.W911NF-07-1-0493 with funds from the DARPA OLE program.

\section{References}

\noindent [1]G. Santarelli, A. Clairon, S. N. Lea, G. M. Tino, Opt. Comm. \textbf{104}, 339 (1994).\\
\noindent [2]H.N. Rutt: J. Phys. E \textbf{17}, 704 (1984).\\
\noindent [3]U. Sch\"{u}nemann, H. Engler, R. Grimm, M. Weidem\"{u}ller, M. Zielonkowski, Rev. Sci. Instrum. \textbf{70}, 242 (1999).\\
\noindent [4]G. Ritt, G. Cennini, C. Geckeler, M. Weitz, Appl. Phys. B \textbf{79}, 363 (2004). \\
\noindent [5]T. Stace, A. Luiten, R. P. Kovacich, Meas. Schi. Technol. \textbf{9}, 1635 (1998).\\
\noindent [6] D. A.  Steck, \emph{ ${}^{87}\mathrm{Rb}$  D Line Data},\\
\noindent  http://steck.us/alkalidata/rubidium87numbers.1.6.pdf.

\end{document}